*Title*

Natural direct effects of vaccines and post-vaccination behaviour

*Authors*

Bronner P. Gonçalves[1,2], Piero L. Olliaro[2], Sheena G. Sullivan[3,4], Benjamin J. Cowling[5]

*Affiliations*

[1] University of Surrey, Guildford, United Kingdom

[2] ISARIC, Pandemic Sciences Institute, University of Oxford, Oxford, United Kingdom

[3] School of Clinical Sciences, Monash University, Melbourne VIC, Australia.

[4] Fielding School of Public Health, University of California, Los Angeles (UCLA), Los Angeles, California

[5] WHO Collaborating Centre for Infectious Disease Epidemiology and Control, School of Public Health, Li Ka Shing Faculty of Medicine, The University of Hong Kong, Hong Kong Special Administrative Region, People's Republic of China

*Correspondence*

bronnergoncalves@gmail.com , piero.olliaro@ndm.ox.ac.uk , bcowling@hku.hk

*Abstract*

Knowledge of the protection afforded by vaccines might, in some circumstances, modify a vaccinated individual's behaviour, potentially increasing exposure to pathogens and hindering effectiveness. Although vaccine studies typically do not explicitly account for this possibility in their analyses, we argue that natural direct effects might represent appropriate causal estimands when an objective is to quantify the effect of vaccination on disease while "blocking" its influence on behaviour. There are, however, complications of a practical nature for the estimation of natural direct effects in this context. Here, we discuss some of these issues, including exposure-outcome and mediator-outcome confounding by healthcare seeking behaviour, and possible approaches to facilitate estimates of these effects. This work highlights the importance of data collection on behaviour, of assessing whether vaccination induces "riskier" behaviour, and of understanding the potential effects of interventions on vaccination that could "turn off" vaccine's influence on behaviour.

*Background*

Vaccination is a preventive intervention whose uptake may be correlated with other health-protective behaviours. However, vaccination may also lead to risk compensation, resulting in additional exposures compared to unvaccinated individuals. Hence, the real-world impact of vaccines could be affected by behavioural changes that occur due to knowledge of the vaccine's protective effect. This risk compensation could result in a reduced level of protection from vaccination in real life compared to what might be expected in a blinded vaccine trial.

Specifically, vaccinated individuals might engage in activities that increase their risk of infection, either by choice or because of societal or cultural norms, e.g., by increasing the number of social contacts that can lead to an infection. Thus, during the COVID-19 pandemic, several studies [1-4] were performed to assess the influence of vaccination on behaviour associated with infection risk. For example, in France, McColl and colleagues [5] analysed online surveys on protective behaviours and showed that vaccinated individuals were less likely to avoid social gatherings and to wear masks in some of the months when surveys were performed. In Canada [6], vaccinated individuals with comorbidities had more social contacts in the third wave of the COVID-19 pandemic compared to unvaccinated individuals with comorbidities. Consistent with this, knowledge about vaccination has been associated with high-risk behaviour [7].

Here, we argue that, although risk compensation might not be pervasive, effectiveness studies should consider this unintended consequence of vaccination both to quantify an effect that is independent of changes in behaviour and to understand potential benefits of interventions that could prevent risk compensation. In particular, we suggest that, along with estimands typically targeted in vaccine studies, a causal estimand that corresponds to the effect of vaccination in the presence of an intervention that blocks its influence on behaviour might be of value for policy makers, in addition to being of scientific interest in its own right. Such an intervention could, for example, involve encouraging individuals not to change behaviour after vaccination, which might be, in some settings, more realistic than interventions that aim to set behaviour to a particular level for all individuals in a population. In fact, this latter type of intervention could be conceived of as being related to controlled direct effects, whilst the former type of intervention (that does not fix behaviour of all individuals to the same level, but rather, by blocking the influence of vaccination, allows natural variation in behaviour) is relevant to natural direct effects.

Below, to facilitate analyses of direct (with respect to behavioural factors) effects of vaccines on clinical outcomes, we present relevant causal diagrams, formally define the proposed effect, which is, as implied above, a natural direct effect, discuss identification, and compare this effect with that estimated in blinded trials. Further, we discuss practical issues in the study of this estimand. For completeness, we also discuss the distinct question of assessing effects of vaccination on behaviour when information on the latter is not available and the potential impact of interference. Throughout, we focus on observational studies and consider confounding by healthcare seeking behaviour, which might affect many vaccine effectiveness analyses.

*Epidemiologic context and notation*

Consider a hypothetical study undertaken during an epidemic. Let $A$ denote vaccination (1 = vaccinated, 0 = not vaccinated), and $Y$, the outcome of interest (1 = disease caused by the pathogen targeted by vaccination, 0 = no disease caused by the pathogen). In *Panel I* of **Figure 1**, we present the scenario described above; $B$ corresponds to infection-related behaviour (e.g., number of social contacts of a relevant type per time unit) and is assessed after vaccine assignment. We assume that $B$ is a mediator of the effect of $A$ on $Y$. Although we consider a scenario where vaccination, on average, leads to "riskier" behaviour among the vaccinated, in some situations, the behavioural changes might be linked to absence of vaccination. For example, compliance with public health measures during the COVID-19 pandemic was reported, in some countries, to be lowest among unvaccinated individuals [8]. Another possibility is for absence of vaccination to require safer behaviours. For example, some jurisdictions or healthcare facilities have required healthcare workers who choose not to be vaccinated against influenza to wear face masks [9, 10].

In observational studies, there might be factors that determine both vaccination and outcome; we denote common causes of $A$ and $Y$ (that is, confounders) by $L$. One of these common causes is healthcare seeking behaviour; for instance, individuals who are more likely to be vaccinated might also be more likely to go to a clinic or hospital if they become sick. In *Panel II* of **Figure**

**1**, we illustrate this with variable $H$ (healthcare seeking behaviour; note the similarity with Figure 1 in [11]). In *Panel III* of **Figure 1**, we assume that $H$ also affects $B$. Below, we consider both *Panels I* and *III* and assume that information on $B$ and $L$ (including $H$) could be collected using, for example, social contact diaries (see [6]) and medical registry databases.

We also define the potential outcomes variables $Y^a$ and $B^a$ as the values that $Y$ and $B$ would take had $A$ been set to $a$. The potential outcomes $Y^{a,b}$ denote the values $Y$ would take under interventions on $A$ and $B$; by the composition assumption, $Y^a = Y^{aB^a}$. The causal estimand that is relevant in real-world settings ($\tau_{rw}$) is thus the following:

$$\tau_{rw} = E[Y^{a=1} - Y^{a=0}] = E[Y^{a=1,B^{a=1}} - Y^{a=0,B^{a=0}}]$$

Note that $\tau_{rw}$ is identified by design in a randomized trial without blinding. As discussed in [12], in trials with blinding, the target effect is different (see also, below, the section *Relation to the estimand in vaccine trials*). Finally, notice that in *Panel III* of **Figure 1**, to better explain the effect proposed below, we present another factor affecting $B$, work-related environment ($W$).

**Figure 1**. Causal diagrams representing relations between vaccination, behaviour, and the outcome of interest. In *Panel I*, we present the causal structure for settings with no confounding; the variables are defined in the main text, and correspond to vaccination ($A$), infection-related behaviour ($B$), and clinical disease ($Y$). In *Panel II*, to illustrate confounding, the variable $H$, healthcare seeking behaviour, is assumed to affect both $A$ and $Y$; we do not present other confounders ($L$). In *Panel III*, we assume that healthcare seeking behaviour affects the exposure ($A$), the mediator ($B$) and the outcome ($Y$); the variable $W$ corresponds to work-related environment. Although it could be argued that in some settings $W$ also directly affects $Y$, here this variable is shown primarily to illustrate causal influences on $B$ in the context of natural direct effects (see section *Natural direct effect of vaccination*). Finally, note that $H$ itself might be influenced by risk aversion; however, this is not shown in **Figure 1** as risk aversion might affect variables in the diagram only through its effect on $H$.

*Panel I*

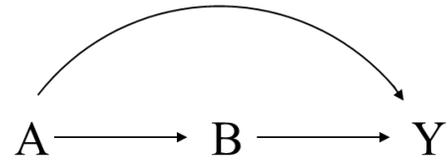

*Panel II*

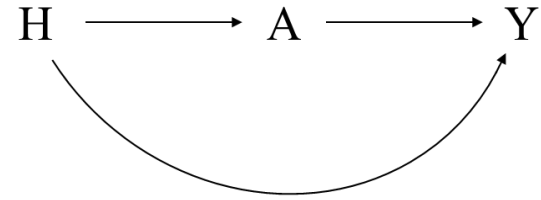

*Panel III*

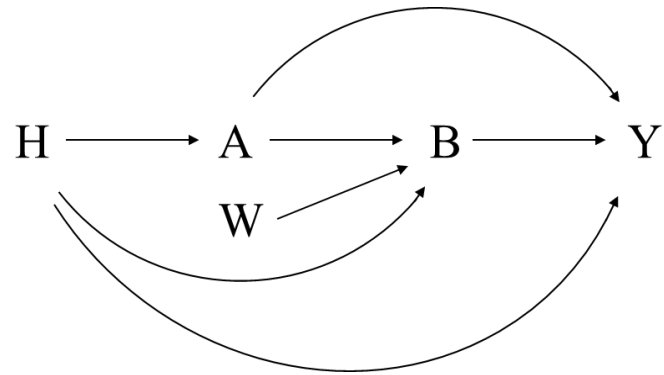

*Natural direct effect of vaccination*

Here, we describe the natural direct effect ($NDE$) of vaccination relative to $B$. This is the effect of $A$ on $Y$ if we were able to "turn off" the path from $A$ to $B$, and is defined as:

$$NDE = E[Y^{a=1,B^{a=0}} - Y^{a=0,B^{a=0}}] = E[Y^{a=1,B^{a=0}} - Y^{a=0}]$$

where $Y^{a=1,B^{a=0}}$ is the value that $Y$ would take had $A$ been set to 1 (vaccination) and $B$ been set to the value it would take absent vaccination, $B^{a=0}$ (which could be different for different individuals). This estimand [13, 14] corresponds to the effect of $A$ on $Y$ when its effect on $B$ is blocked, so that the mediator takes its natural value. This could be represented in **Figure 1** by removing the arrow $A \rightarrow B$ [15]; in [16], this type of effect is represented by edges crossed out of the graph. Omitting this edge would imply that $A$ had no influence on $B$, and $B$ would vary based on other possible causes (e.g., $W$ would still affect $B$). Note that this differs from controlled direct effects as the latter involves fixing the value of the mediator for the entire population (i.e., omitting all edges pointing to the mediator).

In **Table 1**, we present a numerical example to illustrate quantitative differences between $\tau_{rw}$ and $NDE$. In the example, vaccination increases the probability of risky behaviour, and $\tau_{rw}$ has a lower absolute value compared to $NDE$.

**Table 1.** Numerical example. In the table, for simplicity, we consider binary variables. The variable $B$ indicates the event of the number of social contacts being above a pre-defined value, and the outcome, $Y$, represents clinical disease during follow-up. The values for $\Pr(B^{a=1} = 1)$ and $\Pr(B^{a=0} = 1)$ were chosen to illustrate that vaccination can promote risky behaviour.. Note that although $\tau_{rw}$ and $NDE$ were defined on the risk difference scale, vaccine studies typically report effectiveness, which, numerically, represents the risk difference relative to the risk in the unexposed group and here corresponds to $\tau_{rw}^{VE} = 1 - \frac{E[Y^{a=1,B^{a=1}}]}{E[Y^{a=0,B^{a=0}}]}$ and $NDE^{VE} = 1 - \frac{E[Y^{a=1,B^{a=0}}]}{E[Y^{a=0,B^{a=0}}]}$. For the example, the value of $\tau_{rw}^{VE}$ would be (in percentage) 32.5%, and that of $NDE^{VE}$, 42.5%.

| **Behaviour** | |
|---|---|
| | *Proportions* |
| $\Pr(B^{a=1} = 1)$ | 0.70 |
| $\Pr(B^{a=1} = 0)$ | 0.30 |
| $\Pr(B^{a=0} = 1)$ | 0.30 |
| $\Pr(B^{a=0} = 0)$ | 0.70 |

| **Clinical outcome** | |
|---|---|
| | *Proportions* |
| $\Pr(Y^{a=1,b=1} = 1)$ | 0.21 |
| $\Pr(Y^{a=0,b=1} = 1)$ | 0.35 |
| $\Pr(Y^{a=1,b=0} = 1)$ | 0.14 |
| $\Pr(Y^{a=0,b=0} = 1)$ | 0.25 |

| **Total effect** | |
|---|---|
| | *Expected potential outcomes and causal estimand* |
| $\Pr(Y^{a=1} = 1)$ | 0.70 x 0.21 + 0.30 x 0.14 = 0.19 |
| $\Pr(Y^{a=0} = 1)$ | 0.30 x 0.35 + 0.70 x 0.25 = 0.28 |
| $\Pr(Y^{a=1} = 1) - \Pr(Y^{a=0} = 1)$ | -0.09 |

| **NDE** | |
|---|---|
| | *Expected potential outcomes and causal estimand* |
| $\Pr(Y^{a=1,B^{a=0}} = 1)$ | 0.30 x 0.21 + 0.70 x 0.14 = 0.16 |
| $\Pr(Y^{a=0,B^{a=0}} = 1)$ | 0.30 x 0.35 + 0.70 x 0.25 = 0.28 |
| $\Pr(Y^{a=1,B^{a=0}} = 1) - \Pr(Y^{a=0,B^{a=0}} = 1)$ | -0.12 |

*Identification*

Identification of natural direct effects has been extensively discussed in the causal inference literature [13, 17, 18]. In particular, to identify $E[Y^{a=1,B^{a=0}}]$, standard causal assumptions are not sufficient. When $L$ represents the set of common causes of $A$, $B$ and $Y$ (as in *Panel III* of **Figure 1,** with $H$ replaced by $L$), the following assumptions allow nonparametric identification (below, ⫫ indicates independence):

1) $Y^{a,b} ⫫ A|L$
2) $Y^{a,b} ⫫ B|A, L$
3) $B^a ⫫ A|L$
4) $Y^{a,b} ⫫ B^{a^*}|L$

In words, within levels of $L$, potential values of $Y$ and of $B$ are independent of vaccination status. Further, the potential value of the outcome is independent of behaviour, given vaccination and $L$. Assumption 4 is a cross-world independence assumption [19, 20], as it involves potential outcome variables under vaccination levels that might not be compatible. In addition to these, positivity assumptions such as $0 < Pr(A = a|L) < 1$ and $Pr(B = b|L, A = a) > 0$ for all $a,b = 0,1$ need to hold [21]. For *Panel I* of **Figure 1**, Assumptions 1, 3 and 4 would hold unconditionally.

Under these assumptions, we can identify the $NDE$:

$$E[Y^{a=1,B^{a=0}} - Y^{a=0,B^{a=0}}]$$
$$= \sum_{b,l} \{E[Y|A = 1, B = b, L = l] - E[Y|A = 0, B = b, L = l]\} \Pr(B = b|A = 0, L = L) \Pr(L = l)$$

*Expression 1*

The plausibility of these assumptions in observational studies is likely to vary in different contexts. For example, healthcare seeking behaviour ($H$) is a well-described confounder in studies on total vaccine effects. Addressing confounding by this variable requires quantification of $H$ or of proxy variables, and adjustment for healthcare seeking behaviour is often assumed (or not explicitly considered) in effectiveness studies that target $\tau_{rw}$. In the context of natural direct effects, the link $H \rightarrow B$ might be present, and healthcare seeking behaviour might also correspond to an exposure-mediator and mediator-outcome confounder. Indeed, while $H$ might directly affect $Y$ by, for instance, influencing the probability that cases are medically attended, it is possible that $H$ affects $B$, which would imply that $H$ could also affect $Y$ via $B$. Note that although accounting for healthcare seeking behaviour might be difficult, this difficulty is also present in studies on total effects. Finally, other factors (e.g., age) may also bias $\tau_{rw}$ and $NDE$, and data on these factors would be required for vaccine mediation analysis.

*Practical issues*

While the $NDE$ would provide information that is not obtainable from typical estimands in vaccine studies, there are practical issues that need to be considered for this type of effect. First, $B$ might be imperfectly measured. In this case, methods have been described for mediation analyses in the presence of misclassification [22]. For instance, Ogburn and VanderWeele [23], in studying non-differential misclassification of a binary mediator, showed that a $NDE$ estimate with this type of measurement error would fall between the true natural direct effect and the total effect (see also [24]).

Another issue is the possible difficulty in quantifying some aspects of behaviour linked to infection. For example, McColl and colleagues, in studying protective behaviour and SARS-CoV-2 vaccination, analysed data on several types of practices that might influence infection risk [5]; however, some epidemiologic studies might only be able to capture information on one or two aspects of behaviour. Moreover, the underlying causal structure might be more complex. For instance, in **Figure S1** (*Appendix*), we use two behaviour variables, $B_M$ and $B_{SC}$; these variables correspond, respectively, to mask use and to number of contacts. In this case, it might still be possible to identify vaccine effects through specific paths (path-specific effects [16]); see also work by Zhou and Yamamoto [25] on path-specific effects in the presence of multiple, causally-ordered mediators. In the *Appendix*, we discuss path-specific effects for **Figure S1**.

Finally, while capturing information on confounders such as age might be straightforward, accounting for healthcare seeking behaviour might require databases with long-term follow-up (with pre-vaccination period) and that are rich enough to include individual-level data on different aspects of health-related behaviour. There are numerous examples of analyses that used data related to healthcare seeking behaviour during the COVID-19 pandemic [26, 27]. In **Table 2**, we compare data sources for variables $B$ and $H$.

**Table 2.** Data on healthcare seeking behaviour and infection-related behaviour. See also [28], where the authors identified markers for healthcare seeking behaviour in electronic health records, and the work by Bedson and colleagues [29] that argues for disease models integrated with behavioural factors.

|  | **Healthcare seeking behaviour** | **Infection-related behaviour** |
|---|---|---|
| *Examples of variables* | Number of vaccinations targeting another pathogen in the years prior to the study<br><br>Number of diagnostic tests in the year(s) prior to the study<br><br>Participation in medical (e.g. cancer) screening activities<br><br>Factors that might affect healthcare seeking behaviour and that might also be confounders via other paths: age, sex, comorbidities, socioeconomic situation | Mask use (Respiratory infections)<br><br>Number of social contacts (e.g., number of unprotected sexual contacts, for sexually transmitted conditions)<br><br>Factors that might affect infection-related behaviour and that might also be confounders: age, sex, comorbidities, socioeconomic situation |
| *Timing* | Information collected at baseline (before time zero of follow-up) | Data collected after time zero of follow-up and before outcome occurrence |
| *Potential data sources* | Medical registry databases, insurance information | Prospective data collection (e.g. social contact diaries, interviews) |

*An alternative causal structure*

Above, we assumed that healthcare seeking behaviour is a time-invariant (at least with respect to the study time scale) characteristic of individuals. In this section, we discuss an alternative causal structure in which healthcare seeking behaviour might be affected by vaccination and thus vary over time. In **Figure 2**, variables $H_1$ and $H_2$ correspond to healthcare seeking behaviour before and after vaccine assignment, respectively; in this case, the total mediating effect through $B$ includes the paths $A \to B \to Y$ and $A \to H_2 \to B \to Y$. Here, Assumption 4 would not hold, as $H_2$ is a confounder of the mediator-outcome relation affected by $A$. In this situation, although natural direct effects would not be nonparametrically identified based on Assumptions 1 – 4, alternative approaches are available: Imai and Yamamoto, using linear models, describe sensitivity analyses for this scenario [30]; Tchetgen Tchetgen and VanderWeele [31] discuss other identification assumptions; bounds that do not require Assumption 4 could also be used [19]; Díaz and Hejazi [32] showed that direct effects for stochastic interventions require weaker assumptions; finally, see work by VanderWeele and colleagues on three methods (including one method in which the confounder affected by the exposure and the mediator are considered jointly) [33].

**Figure 2.** Alternative causal structure representing relations between vaccination ($A$), infection-related behaviour ($B$), healthcare seeking behaviour ($H_1$ and $H_2$) and clinical outcomes ($Y$). Here, vaccination affects healthcare seeking behaviour, and thus two variables, for two time points (before, $H_1$, and after, $H_2$, vaccination), are presented. Note that in some settings, we might want to consider pre-vaccination infection history as a variable that is a parent node of some of the variables in this diagram, for instance $A$, $B$ and Y, and in this case adjustment for such a variable would be necessary.

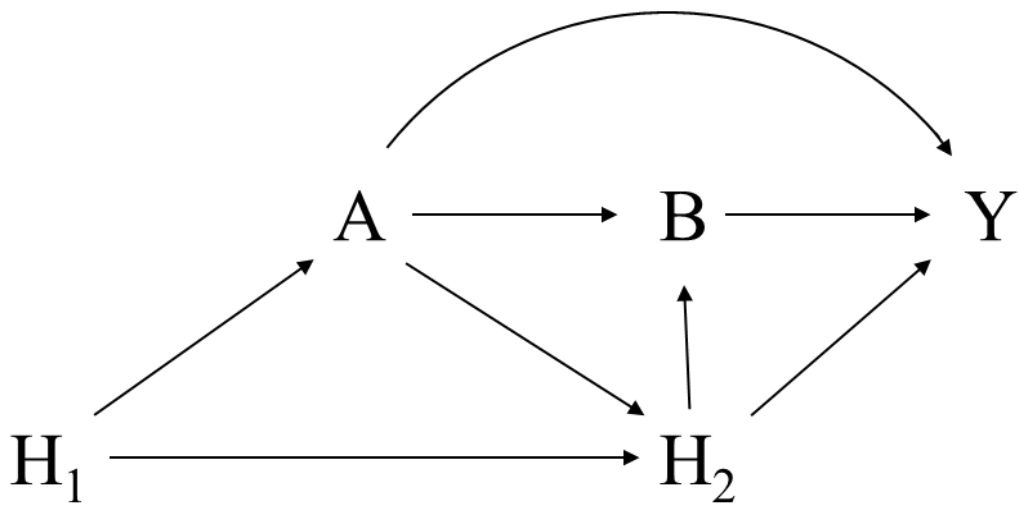

*Relation to the estimand in vaccine trials*

In a trial with blinding and in which the vaccine being tested does not lead to clear side-effects (as the latter would effectively reveal treatment assignment for a fraction of the participants), vaccination status is not known and hence cannot affect perception of protection. Thus, the estimand in trials might be different from $\tau_{rw}$ and $NDE$; see also [12], where this is discussed under different causal assumptions.

To define the estimand in blinded trials, we consider an additional variable $P$, perception of protection, and assume that the effect of $A$ on $B$ is entirely mediated by $P$ (**Figure S2**). We also assume: a) that $P$ takes values from the set $\{-1, 0, 1\}$; b) that in settings without blinding, $P = A$; and c) that in blinded trials, $P$ is set to $-1$, which denotes uncertainty about protection. Given this, in a vaccine trial, vaccination effect on $Y$ ($\tau_{vt}$) is:

$$\tau_{vt} = E[Y^{a=1,p=-1} - Y^{a=0,p=-1}]$$
$$= E[Y^{a=1,p=-1,B^{a=1,p=-1}} - Y^{a=0,p=-1,B^{a=0,p=-1}}]$$
$$= E[Y^{a=1,B^{p=-1}} - Y^{a=0,B^{p=-1}}]$$

where the second equality holds because $Y^{a,p} = Y^{a,p,B^{a,p}}$ and the third follows from $Y^{a,p,b} = Y^{a,b}$, as implied by **Figure S2** (see for example Section 7.3 in [34]). So, under the causal diagram in **Figure S2**, a difference between $\tau_{vt}$ and the other estimands ($\tau_{rw}$ and $NDE$) is that in a blinded trial, $B$ might take a different value due to uncertainty about vaccine-related protection.

*Use of alternative outcomes to investigate unascertained vaccine effects on behaviour*

Many vaccine studies do not collect information on behaviour, which prevents direct assessments of $A \to B$. Here, we consider scenarios where we might learn about vaccine effects on behaviour despite lack of data on behavioural changes.

In *Panel I* of **Figure 3**, we show a causal structure in which an outcome, $R$, other than the outcome of interest ($Y$), is presented; $R$ corresponds to infections caused by other pathogens. For instance, $Y$ could represent COVID-19-related hospitalization and $R$, hospitalization by another respiratory pathogen. We assume that aspects of behaviour relevant to $Y$ are also relevant to $R$, and then ask: *When could data on R provide information on A → B (if data on B are not available)?* Under causal assumptions in *Panel I* of **Figure 3** (e.g., absence of cross-reactive immunity $A \to R$), and after conditioning on $H$, an association between $A$ and $R$ would indicate that vaccination affects behaviour. In *Panel II*, some aspects of behaviour are only relevant for the $Y$-causing pathogen; here, a vaccine effect on $R$ would also suggest risk compensation. In *Panel III*, however, some components of behaviour are only relevant for $R$, and thus learning that $A$ affects $R$ does not imply vaccine effects on behaviour relevant for $Y$. Note also that, in *Panel I*, if $Y$ affected $R$ (e.g., if individuals with the $Y$-causing pathogen are [temporarily] resistant to the $R$-causing pathogen), then $A$ and $R$ would not be d-separated, when controlling for $H$, if the edge $A \to B$ was not present, as there would exist the path $A \to Y \to R$ (further, conditioning on $Y$ would open the path $A \to Y \leftarrow B \to R$).

Note that recent work [35] on hidden mediation analysis could be used to quantify direct and indirect effects of vaccination with respect to behaviour; in this case, $R$ would be a proxy variable of behaviour (see section 3 of [35]); see also [36] on a structurally related question. It is also important to mention work on negative control populations [37]; in our context, if, in a subgroup of the population, immune responses are unaffected by vaccination (e.g., due to immunosuppression), an association between $A$ and $Y$ in this subgroup would suggest unmeasured confounding or vaccine effects through other paths, including possibly via behaviour. Finally, and importantly, Stensrud and colleagues [12], in discussing their results, under a different causal structure, proposed using negative controls to assess the impact of vaccination on behaviour.

**Figure 3**. Causal diagrams illustrating vaccine effects in the presence of an alternative infection outcome. In *Panels I-III*, nodes $A, B, Y, R$ correspond, respectively, to vaccination, infection-related behaviour, clinical outcome linked to the pathogen of interest, and an alternative infection outcome (caused by another pathogen, not targeted by vaccination). In *Panel I*, $H$ denotes healthcare seeking behaviour. In *Panels II* and *III*, $B_1$ and $B_2$ represent different aspects of behaviour that might affect $Y$ and $R$. Note that confounders of the relation between $B$ and $R$ (e.g., $C_2$; not shown in the figure) would not imply additional non-causal open paths between $A$ and $R$, as $B$ would be a collider in the path $A \to B \leftarrow C_2 \to R$ (see, for example, section 5.2 in [38]).

*Panel I*

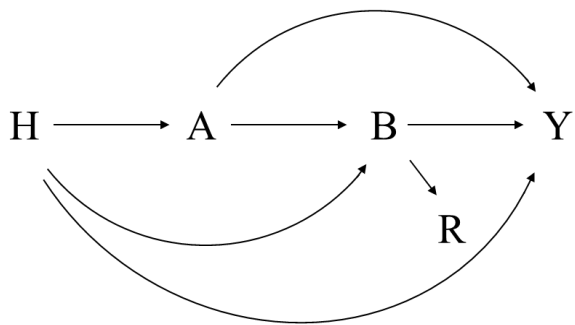

*Panel II*

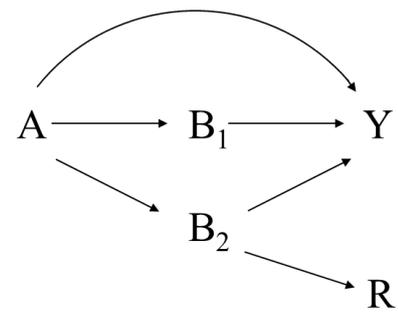

*Panel III*

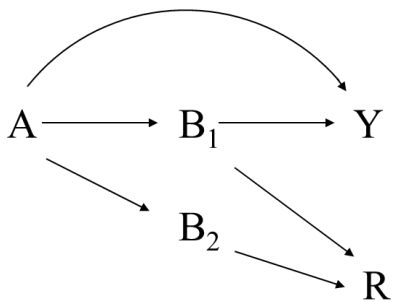

*The potential impact of interference*

In our discussion above, we assume for simplicity no interference, i.e., the vaccination status of an individual does not affect the outcome of other individuals. However, although this "no-interference" assumption is often made in vaccine studies [39-41], interference is likely in many studies on infectious diseases. In fact, even for non-communicable conditions, dependence linked to social networks can result in spurious results [42].

In the presence of interference, potential outcomes are appropriately expressed using vectors of vaccine assignment to (subgroups of) the population. Using definitions of variables above and assuming that the population is partitioned into $n$ subgroups, then potential outcomes of individual $j$ in group $i$ (with $i = 1, ..., n$ and $j = 1, ..., n_i$) correspond to $Y_{i,j}^{a_i}$, where $a_i$ is a vaccine assignment vector in group $i$; by expressing dependence of potential outcomes on group-specific assignment, we make a partial interference assumption [43].

When interference is present, various effects might be defined. For instance, individual total causal effects correspond to $Y_{i,j}^{a_{i\setminus j}, a_j=1} - Y_{i,j}^{a_{i\setminus j}^*, a_j=0}$, where $Y_{i,j}^{a_{i\setminus j}, a_j}$ is the potential outcome of individual $j$ in group $i$ where individual $j$ is assigned $a_j$ and assignment to all other individuals in group $i$ is represented by $a_{i\setminus j}$. Another effect that might be studied under interference is the following direct effect ("direct" in reference to vaccine assignment of other individuals; see section 6 in [43] for discussion on direct effect terminology in mediation and interference): $Y_{i,j}^{a_{i\setminus j}, a_j=1} - Y_{i,j}^{a_{i\setminus j}, a_j=0}$. Group- and population-level versions of this effect can also be defined [43]. In the context that we consider in this paper, the following individual-level effect might be defined:

$$NDE_{i,j\,(interf, a_{i\setminus j})} = Y_{i,j}^{a_{i\setminus j}, a_j=1, B_{i,j}^{a_j=0}} - Y_{i,j}^{a_{i\setminus j}, a_j=0}$$

where $Y_{i,j}^{a_{i\setminus j}, a_j=1, B_{i,j}^{a_j=0}}$ is the potential outcome of individual $j$ in group $i$ where individual $j$ is assigned vaccination status 1 but her behaviour variable takes the value it would have taken in absence of vaccination, $B_{i,j}^{a_j=0}$, and assignment for all other individuals in group $i$ is represented by $a_{i\setminus j}$. Note that this notation assumes that behaviour is not affected by interference; however, it is conceivable that it might also depend on the group-level assignment vector, $B_{i,j}^{a_{i\setminus j}, a_j=0}$.

In observational studies with interference, identification and estimation are, in general, not straightforward. There have been important developments in this area [43-45], and although beyond the scope of the current work, future research on natural direct effects of vaccines could extend this existing literature, including on causal diagrams for interference (see subsection 2.2, which includes discussion on path-specific effects under interference, and sections 3 and 4 in [46]) to clarify identification conditions under interference, or on simulation approaches for networks [47] to assess its potential impact on causal inferences. Note that in contexts with interference, the separate question of whether vaccination has spillover effects (that is, $Y_{i,j}^{a_{i\setminus j}, a_j=0} - Y_{i,j}^{a_{i\setminus j}^*, a_j=0}$) is important, and quantification of this type of effect's component that is related to other individuals' behaviour warrants further investigation in future studies.

*Discussion*

Risk compensation might affect vaccine effectiveness, typically if vaccinees would behave differently from unvaccinated people when social distancing and other preventative measures are adopted during epidemics, as it has been shown during the COVID-19 pandemic [48, 49]. Thus, it might be particularly important to consider the *NDE* in these scenarios, even though its identification requires additional assumptions compared to $\tau_{rw}$ and to controlled direct effects. When estimation of *NDE* is feasible (i.e., where data on behaviour were collected and identification assumptions are likely to hold), *NDE* would provide information on vaccine effect if risk compensation could be avoided, hence help inform health policy decisions to invest in strategies to minimise the impact of vaccination on behaviour.

This paper highlights the implications of post-vaccination infection-related behaviour *versus* healthcare seeking behaviour when designing empirical studies and the need to collect information on behaviour and to investigate vaccine's natural direct effects. In trials where blinding prevents participants from knowing vaccination status, vaccine effects on behaviour are less likely – thus the estimated vaccine effect might not correspond to the total effect in real-world settings, where protection depends on immune responses plus behaviour. A recent study [12] formally described the difference between the causal effect targeted in vaccine trials and the effect which operates in communities after vaccine deployment, proposing definitions of vaccine effects accounting for participants' beliefs on vaccination status. The estimands proposed provide information on both immunological and behavioural effects of vaccination by defining interventions that correspond to messages with possible vaccination status (see expressions 3 and 4 in [12]). The type of effect described here is different: we focus on studies that collect behavioural information; and we make a distinction between controlled and natural direct effects. We believe that, in some settings, interventions implied by natural direct effects of vaccines are more plausible than those related to controlled direct effects; note however that we do not claim that the latter type of effects is not useful for vaccine-related questions – rather, the relative usefulness of natural versus controlled direct effects might depend on the type of behaviour being considered, which determines the intervention's real-world acceptability. On a related issue, it is possible that participation in trials might be influenced by healthcare seeking behaviour, and that might further affect the external validity [50] of these studies.

We should mention that although there is evidence for changes in protective behaviour post-vaccination, some studies reported no behavioural changes after vaccination [3, 51] – which could be linked to cultural factors, or, suggest efficient communication by health authorities. We also know that introducing preventative measures could have effects on behaviour that extend beyond the condition being targeted: for instance, it has been reported that initiation of HIV pre-exposure prophylaxis might affect sexual behaviour [52], exposing to increased risk of sexually transmitted infections.

The picture presented here is a simplification of reality, although similar simplifications are often used in epidemiologic studies on vaccines. Indeed, it is possible that future studies might find that the impact of behavioural changes after vaccination varies considerably in different settings. Consistent with this, using dynamical modelling of disease transmission, Shaw and Schwartz showed in the context of adaptive networks (that is, networks where links between individuals can be "rewired") that considerably fewer vaccine resources were needed for epidemic extinction compared to static networks [53]. Moreover, during an epidemic, there might exist temporal trends in protective behaviour, that might follow different patterns for different types of behaviours [54]; analyses that aim to quantify natural direct effects of vaccines need to account for this. Further note that while we focused on post-vaccination behaviour, other issues, such as changes in pathogen population composition over time and waning immunity also play a role but are broadly relevant to vaccine studies and were extensively studies for COVID-19 vaccines.

**Source of funding**

This work was supported by the UK Foreign, Commonwealth and Development Office and Wellcome (215091/Z/18/Z, 222410/Z/21/Z, 225288/Z/22/Z, and 220757/Z/20/Z); the Bill & Melinda Gates Foundation (OPP1209135); and the philanthropic support of the donors to the University of Oxford's COVID-19 Research Response Fund (0009109). This work was also supported by an RGC Senior Research Fellowship from the University Grants Committee of Hong Kong to BJC (grant number: HKU SRFS2021-7S03).

**Conflict of interest statement**

BJC consults for AstraZeneca, Fosun Pharma, GlaxoSmithKline, Haleon, Moderna, Novavax, Pfizer, Roche, and Sanofi Pasteur. All other authors declare no competing interests.

*Appendix*

Table of contents

*Path-specific effects and infection-related behaviour*

*Figure S1*

*Figure S2*

*Reference*

*Path-specific effects and infection-related behaviour*

Here, we briefly discuss **Figure S1**, where two different aspects of behaviour are presented: $B_M$ corresponds to mask use, and $B_{SC}$, to number of social contacts. In this setting, we might be interested in the effect of vaccination on the outcome if its effect on mask use, but not that on social contacts, were "turned off". This could be defined as:

$$E\left[Y^{a=1,B_{SC}^{a=1},B_M^{a=0}} - Y^{a=0,B_{SC}^{a=0},B_M^{a=0}}\right]$$

Similarly, the effect of vaccination on the clinical outcome if we were able to block its effect on the number of social contacts would be:

$$E\left[Y^{a=1,B_{SC}^{a=0},B_M^{a=1}} - Y^{a=0,B_{SC}^{a=0},B_M^{a=0}}\right]$$

Further, based on **Figure S1**, the total effect can be decomposed (see also [1]):

$$E\left[Y^{a=1,B_{SC}^{a=1},B_M^{a=1}} - Y^{a=0,B_{SC}^{a=0},B_M^{a=0}}\right]$$
$$= E\left[Y^{a=1,B_{SC}^{a=1},B_M^{a=1}} - Y^{a=1,B_{SC}^{a=1},B_M^{a=0}}\right] + E\left[Y^{a=0,B_{SC}^{a=1},B_M^{a=0}} - Y^{a=0,B_{SC}^{a=0},B_M^{a=0}}\right]$$
$$+ E\left[Y^{a=1,B_{SC}^{a=1},B_M^{a=0}} - Y^{a=0,B_{SC}^{a=1},B_M^{a=0}}\right]$$

where the first two average contrasts correspond to mediated effects and the third, to a direct effect.

**Figure S1.** Causal structure including two aspects of behaviour potentially affected by vaccination. $B_M$ corresponds to mask use, and $B_{SC}$, to number of social contacts.

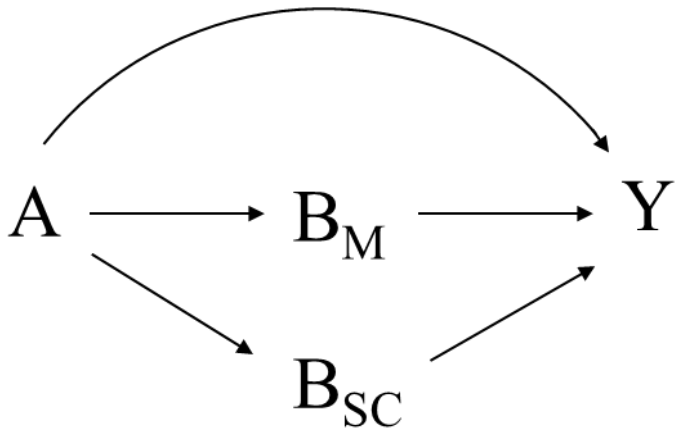

**Figure S2.** Causal diagram similar to *Panel I* of **Figure 1** and that also includes the variable *P*, which denotes perception of vaccine-related protection. Note that the causal assumptions implied by this directed acyclic graph include: *A* affects *B* only through *P*; *P* affects *Y* only through *B*.

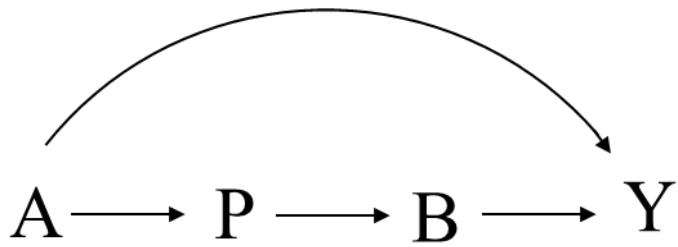